\documentclass[a4paper,12pt]{article}

\usepackage[utf8]{inputenc}
\usepackage[T1]{fontenc} 
\usepackage{authblk}
\usepackage{amsmath}
\usepackage{amsfonts}
\usepackage{graphicx}
\usepackage{amssymb}
\providecommand{\keywords}[1]
{
  \small	
  \textbf{\textit{Keywords---}} #1
}
\title{\boldmath Interacting vacuum at infinity}

 \author{G. Kittou\footnote{georgia.kittou@aum.edu.kw}}
 \affil{Department of Mathematics, \\College of Engineering and Technology, \\American University of the Middle East,\\ Kuwait}

\begin{document}
\maketitle
\abstract{We apply the central extension technique of Poincar\'{e} to dynamics involving an interacting mixture of pressureless matter and vacuum near a finite-time singularity. We show that the only attractor solution on the circle of  infinity  is the one describing a vanishing matter-vacuum model at early times.}

 
\textbf{AMSC:} 83C75, 83F05

\section{Introduction}
\par\noindent
We say that a general relativistic spacetime  exists globally  if it is geodesically complete for an infinite duration of time. On the contrary, singular universes imply the occurrence of a finite-time singularity \cite{YC1,YC2}. Consequently the spacetime is not extendable for eternity. Since Einstein's General Theory of Relativity predicts the existence of finite-time singularities in  cosmological models \cite{HAW1,HAW2}, it is essential to find appropriate tools to analyse the stability of those spacetime singularities. 
\par
A most known method to represent the structure of spacetime at infinity is the method of central projections \cite{Perko, Meiss, Dumortier}. Wherein the structure of a physical spacetime is conformally changed in a way that infinity becomes the boundary of a new unphysical spacetime \cite{skot1}.
\par
In previous works \cite{p1,p2,PhD} and authors in \cite{od3} have studied the asymptotic dynamics near finite-time singularities of flat and curved universes filled with interacting fluids. Especially in \cite{p3}, we considered the GCG proposal to model a unified dark energy model describing an interaction between dark matter and vacuum, and came up with interesting dominant features on approach to a finite-time singularity. 
\par
To be more precise, two asymptotic  solutions  were discussed in details  in \cite{p3}; The first  behaviour was asymptotically obtained in the absence of interaction and in the frame of the GCG model \cite{CHA}. For this case, the dominant features of the  cosmological  model were found to be identical to those of a CDM model \cite{Bento}, with a collapse singularity placed at early times. The second asymptotic behaviour found  describes the case a cosmological model in a Quasi-de Sitter regime  with energy being transferred from dark matter to vacuum. In fact, if the energy transfer from dark matter to vacuum is without bond, the model approaches de Sitter universe with a sudden-type singularity \cite{PhD} placed at late times. 
\par
In this work we are interested in characterising the stability of the aforementioned solutions at infinity using the method of central projections. It is expected that the results of applying the method will give extra information on the qualitative behaviour of the asymptotic orbits in the vicinity of a finite-time singularity. Therefore, we strongly believe that the method of central projections along with the method of asymptotic splittings  will provide a complete description of the dominant part of the vector field at infinity. 
\par 
The plan of this paper is as follows: In Section \ref{sc1}, we define the autonomous system  that  describes the energy exchange between dark matter and vacuum in the GCG regime as found in \cite{p3} and briefly study its local dynamics. In  Section \ref{sc2}, a transformation of variables is introduced  in order to transform the vector field describing the energy exchange. To continue with, on the transformed system,  we apply the method of central projection. In Section \ref{infinity}, we present the qualitative behaviour of the solution at infinity. In Section \ref{cla}, we provide a brief description of the dynamical character of the finite-time singularities using the Bel-Robinson energy. In the last section, we discuss that various results extracted from this paper.
\section{Interacting Vacuum}\label{sc1}     
\par\noindent
Consider a 2-fluid cosmological model in a flat FRW universe describing an energy flow from dark matter to vacuum. The asymptotic dynamics of such a model where exploited in details in \cite{p3}, wherein the Einstein equations for such an interaction read
\begin{equation}\label{e1}
3H^2=\rho_m+\rho_V
\end{equation}
\par\noindent
where $\rho_m$ and $\rho_V$ the energy densities for matter and vacuum respectively and $H$ the Hubble parameter. Assuming an interaction of the form \cite{Wands}
\begin{equation}
   Q=3\alpha H\frac{\rho_m\rho_V}{\rho_m+\rho_V},
\end{equation}
\par\noindent
where $0<\alpha\leq 1$ a positive constant in the GCG regime, the continuity equations read
\begin{eqnarray}
\label{e2}\dot{\rho_m}+3H\rho_m&=&-Q\\
\label{e3}\dot{\rho_V}&=&Q.
\end{eqnarray}
\par\noindent
By combining Eqs. (\ref{e1}), (\ref{e2}) and (\ref{e3}) we end up with the following nonlinear, second order differential equation
\begin{equation}\label{master}
    \ddot{H}+3(\alpha+1)H\dot{H}+2\alpha \dot{H}^2/H=0. 
\end{equation}
\par\noindent
It is very useful to our asymptotic analysis to write the differential equation above in a form of an autonomous dynamical system. By renaming $x=H$ and $y=\dot{H}$ we find 
\begin{eqnarray}\label{system}
\dot{x}&=&y\nonumber\\
\dot{y}&=&-3(\alpha+1)xy-2\alpha\frac{y^2}{x}.
\end{eqnarray}
\par\noindent
We have argued in previous work \cite{p3}, that assuming power-law solutions of the form $H=p/t$, where $p$ a constant, then the system above admits the following solutions
\begin{equation}
    p(2-3p)=0,\quad 0<\alpha\leq 1.
\end{equation}
\par\noindent
For $p=0$ the Hubble parameter is identically equal to zero. Since we are assuming expanding universes, this solution is rejected. On the other hand, for $p=2/3$, the Hubble parameter and the scale factor satisfy
\begin{equation}
    H(t)=\frac{2}{3}t^{-1},\quad a(t)=t^{2/3}.
\end{equation}
\par\noindent
Hence the system admits only one exact power-law solution that describes a matter-dominated universe.
\par
The goal of this section now is to find a qualitative description of the Hubble parameter $x=H$ with respect to time. As shown above, exact solutions may be obtained for  Eq. (\ref{master}). It is more useful though to obtain a complete understanding of the solution space. 
\subsection{Local Analysis}
\par\noindent
 For the nonlinear system above, important information for its global dynamics can be derived by examining the local dynamics of the corresponding linear system (to be determined below), in the neighbourhood  of all possible equilibria. We  note here that  we consider only the cases of expanding universes (that is $H>0$), hence $x>0$. Therefore, the physical phase space is $\mathcal{M}=\{(x,y):x>0, y\in \mathcal{R}\}$. 
 \par
 To begin with the phase space analysis, one must determine the critical points $(x^*, y^*)$ for the system (\ref{system}). These are the points where the time derivatives of the variables $(x,y)$ in Eq. (\ref{system}) vanish simultaneously. Additionally, the equilibrium points  are met at the intersection of the nullclines $N_x$ and $N_y$ \cite{Meiss} as shown below,
\begin{equation}\label{condition1}
    \dot{x}=N_x=0,\quad \dot{y}=N_y=0,
\end{equation}
\par\noindent
and represent the asymptotic behaviour either at the beginning or at the end of the evolution. 
\par\noindent
It occurs that the nullclines are described by the equations below,
\begin{equation}\label{nullclines}
      N_x=\{y=0\}, \quad N_y=\{y=0\}\cup\Big\{y=-\frac{3(\alpha+1)}{2\alpha}x^2\Big\},\quad\textit{for}\quad x>0.
\end{equation}
\par\noindent
Therefore,  for the given system (\ref{system}) there are not any isolated equilibria. Instead, $y^*=0$  consists of a line of equilibria. \par
A comment about the nullclines  of the system is in order. The nullclines $N_x$ and $N_y$ divide the phase space into three regions; For the first region $[x>0, y>0]$, all initial conditions that set off in the first quadrant, eventually reach the line of equilibria $y^*=0$. For the second region $[x>0, -3(\alpha+1)x^2/(2\alpha)<y<0]$, initial conditions move upwards and end up approaching  the line of equilibria $y^*=0$, whereas initial conditions that  satisfy $[x>0, y<-3(\alpha+1)x^2/(2\alpha)]$ eventually move away towards infinity.
\par
We can conclude that the line of equilibria $y^*=0$ (for $x>0$), is a global attractor for all initial conditions in the interior of the first two regions as described above. In the next paragraph, we will affirm the result of this qualitative analysis by linearization and by examining the stability of the line of equilibria. 
\subsection{Stability of critical points in the finite plane}
\par\noindent
An important part of phase plane analysis is to determine the critical points of the system (\ref{system}). These fixed points are met at the intersections of the nullclines, in our case this is the line of equilibria $y^*=0$. The linear part of the system on the line of equilibria $(x^*,0)$ reads
\begin{equation}
\mathcal{D}f(x^*,0)=
  \left(\begin{array}{cc}
    0&1 \\
  0&-3(\alpha+1)x^*
  \end{array}\right).
  \end{equation}
  \par\noindent
The corresponding eigenvalues are
\begin{equation}
    \lambda_1=0\quad\lambda_2=-3(\alpha+1)x^*,
\end{equation}
\par\noindent
with  a pair of correlated   eigenvectors 
\begin{equation}
  \vec{v_1}^T=(1,0)\quad\vec{v_2}^T=[1,-3(\alpha+1)x^*].
\end{equation}
\par\noindent 
Since $x>0$, the line of equilibria $y^*=0$ attracts all other solutions along parallel lines to $\vec{v_2}$ \cite{Meiss}, as already discussed above. Given that  $x=H$ and $y=\dot{H}$, we see that the line of equilibria corresponds to solutions of type $H\propto \mathcal{C}$ and $a(t)\propto \exp{\mathcal{C}t}$, where $\mathcal{C}$ a cosmological constant. Therefore, at late times, under certain initial conditions, trajectories move towards de-Sitter universe. 
\par\noindent
Now, in order to fully describe the qualitative evolution of the system, we must study the asymptotic behaviour of the system around finite-time singularities (if any). This analysis is shown in the next section.
\section{Asymptotic analysis and infinity}\label{sc2}
\par\noindent
It has been argued in \cite{p3} that  in the vicinity of a finite-time singularity, two types of asymptotic solutions arise depending on the value of the GCG constant $\alpha$.  These solutions describe in turn two different epochs in the evolution of the universe; the first solution describes a CDM universe with finite-time singularity placed at early times (see Eq. (32) and Eq. (51) in \cite{p3}). The second solution found (see Eq. (43) in \cite{p3}) describes an exponential expansion of the universe  at late times, and the universe asymptotically approaches the known singular free de-Sitter universe. This is precisely the solution described above in the finite phase plane analysis. 
\par
We note here that in \cite{p3} an intermediate phase of expansion was found when  certain conditions were met with a finite-time singularity placed at late times. Our task now is to study whether or not the solution describing a CDM universe is asymptotically an attractor solution around the finite-time singularity at early times. 
\par
We recall from \cite{p3} that the system   (\ref{system}) consists of the following quasi-homogeneous decompositions \footnote{We say that  a system $(\dot{x},\dot{y})=[P(x,y),Q(x,y)]$ is quasi-homogeneous or weight-homogeneous if there exist $\mathbf{w}=(w_1,w_2)\in\mathcal{N}^2$ and $d\in\mathcal{N}$ such that for arbitrary $\lambda\in\mathcal{R^+}$ we have $P(\lambda^{w_1}x,\lambda_{w_2}y) = \lambda^{w_1-1+d}P(x, y), Q(\lambda^{w_1}x,\lambda^{w_2}y) = \lambda^{w_1-1+d}Q(x, y)$. We call $\mathbf{w}=(w_1,w_2)$ the weight-vector of system $(\dot{x},\dot{y})$ and $d$ the weight degree.}
\begin{eqnarray}
    \label{dc1}f_{I}(x,y)&=&\{[y,-3(\alpha+1)xy]+(0,-2\alpha\frac{y^2}{x})\},\\
    \label{dc2}f_{II}(x,y)&=&\{(y,-2\alpha\frac{y^2}{x})+[0,-3(\alpha+1)xy]\},\\
    \label{dc3}f_{III}(x,y)&=&[y,3(\alpha+1)-2\alpha\frac{y^2}{x}],
    \end{eqnarray}
\par\noindent
Each one of the decompositions above is spitted uniquely  into a dominant and a subdominant vector.  As follows, each dominant part  defines a  unique dominant balance which in turn  describes the evolution of the scale factor in the vicinity of the finite-time singularity. These balances are defined below
    \begin{eqnarray}
  \label{b1}\mathcal{B}_{I}&=&\left[\left(\frac{2}{3(\alpha+1)},-\frac{2}{3(\alpha+1)}\right),(-1,-2)\right]\\
   \label{b2} \mathcal{B}_{II}&=&\left[\left(\theta, \frac{\theta}{2\alpha+1}\right),\left(\frac{1}{2\alpha+1},\frac{-2\alpha}{2\alpha+1}\right)\right],\\
 \label{b3} \mathcal{B}_{III}&=&\left[\left(\frac{2}{3},-\frac{2}{3}\right),(-1,-2)\right],
\end{eqnarray}
\par\noindent
 and are valid asymptotically as $t\rightarrow0$. We note here that each balance above is obtained by substituting in the dominant part of each decomposition above the forms $x(t)=\theta t^(w_1), y(t)=\xi t^(w_2)$ where $(\theta, \xi)\in\mathcal{C}$ and $(p,q)\in\mathcal{Q}$. The dominant exponents $(w_1,w_2)$ define the weight-vector of each balance. As argued before, the second balance (\ref{b2}) describes  the evolution of the universe at late times where all solutions approach asymptotically the de-Sitter universe. 
 \par
 Our goal now is to examine the qualitative behaviour at infinity of the first and third balance, described by the Eq. (\ref{b1}) and
 Eq. (\ref{b3}) respectively. To do so, two  transformation of variables are  applied in the asymptotic system (\ref{system}). The first transformation of variables is applied the next paragraph, wherein the system (\ref{system}) through a change of variables becomes polynomial. A second transformation is then applied in subsection \ref{poincares} the new polynomial system  so that  infinity is  mapped  to a finite point on the so called 'circle of infinity'.  This mapping is known as  Poincar\'{e} compactification \cite{Meiss}. 
 \subsection{The polynomial system}\label{scc3}
In what follows, we apply the  change of variables \cite{transf1, transf2}
\begin{equation}\label{change}
x=u^{1/|w_2|},\quad y=(uv)^{1/|w_1|}
\end{equation}
\par\noindent
with inverse
\begin{equation}\label{tran}
u=x^{|w_2|}\quad v=\frac{y^{|w_1|}}{x^{|w_2|}},
\end{equation}
\par\noindent
and a rescaling of time $dn/dt=u^{1/2}$. This is done in order  to  transform the dominant parts of the quasi-homogeneous decompositions above  into a polynomial system of the form
\begin{equation}\label{transfm}
u'=uf(v),\quad v'=g(v),
\end{equation}
\par\noindent
where prime $(')$ denotes differentiation with respect to the new variable $n$. This change of variables is necessary to be done as Poincar\'{e} compactification is applied only to polynomial vector fields \cite{Perko}. The phase space for the new transformed system (\ref{transfm}) is $\mathcal{\tilde{M}}=\{(u,v):u>0, v\in \mathcal{R}\}$.
\par\noindent
We note here that we apply the the transformation only on  the  ``All-terms-dominant decomposition'' of the system, namely 
Eq. (\ref{dc3}) for $0<\alpha\leq1$\footnote{The first decomposition can be derived by setting $\alpha\rightarrow0$ while the second decomposition is qualitatively described in the local phase portrait analysis in previous sections.}. To continue with, we use the weight-vector $\mathbf{w}=(|w_1|,|w_2|)=(1,2)$ and derive the following polynomial system
\begin{eqnarray}\label{asym1}
u'&=&2uv\nonumber\\
v'&=&-3(\alpha+1)v-2(\alpha+1)v^2.
\end{eqnarray}
\par\noindent
We mention that phase portrait analysis of the transformed system above is qualitatively the same as the local analysis of the original system (\ref{system}). It can be shown that $v^*=0$ is a line of equilibria that asymptotically attracts all solutions under certain initial conditions. Additionally, solutions that set off in the region $(u>0,y<-3/2)$ move away towards infinity.  This is an important result as it shows that the transformation that we use preserves the local properties of the original system (\ref{system}).
In what follows, we will apply the method of central projection in  the  system (\ref{asym1}).
\subsection{Poincar\'{e} compactification}\label{poincares}
\par\noindent
Let us now start our analysis by applying  the transformation $(\ref{change})$ on the decomposition described by the vector $[y,-3(\alpha+1)xy-2\alpha y^2/x]$.  Recall that the dominant properties of the system are described by the balance
\begin{equation}
\mathcal{B}_{III}=\left[\left(\frac{2}{3},-\frac{2}{3}\right),(-1,-2)\right],
\end{equation}
\par\noindent
with a general asymptotic solution \cite{p3} that reads
\begin{equation}\label{sol1}
    x(t)=\frac{2}{3}t^{-1}+c_{31}t^2+\cdots,\quad as\quad t\rightarrow0,\quad \alpha\rightarrow0.
\end{equation}
\par\noindent
In what follows, we discuss the stability of the system (\ref{asym1}) near  finite-time singularities and argue whether or not the asymptotic solution (\ref{sol1}) is an attractor solution at early times. This is accomplished by a projection from $\mathcal{R}^2$ to the northern hemisphere of a sphere (the Poincar\`{e} sphere) \cite{Meiss} described by
\begin{equation}
\mathcal{S}^{2+}=\{(X,Y,Z):X^2+Y^2+Z^2=1,Z\geq0\}.
\end{equation}
\par\noindent 
Geometrically speaking, each point $(u,v)\in\mathcal{R}^2$ is projected, through the center of the sphere, to a unique point in $\mathcal{S}^{2+}$ in such a way  that ``infinity''  for $(u,v)$ is  now mapped to the  circle of infinity $X^2+Y^2=1$. 
\par
Under the map
\begin{equation}\label{map1}
 u=\frac{X}{Z},\quad\text{and}\quad v=\frac{Y}{Z}   
\end{equation}
\par\noindent
the planar system (\ref{asym1}) becomes the singular system \cite{Perko, Meiss, Dumortier,Dover}
\begin{eqnarray}\label{singular}
\dot{X}&=&Z[(1-X^2)P-XYQ]\nonumber\\
\dot{Y}&=&Z[-XYP+(1-Y^2)Q]\nonumber\\
\dot{Z}&=&-Z^2(XP+YQ),
\end{eqnarray}
\par\noindent
where
\begin{eqnarray}
\nonumber P&=&P\left(\frac{X}{Z},\frac{Y}{Z}\right)=\frac{2XY}{Z^2},\\ Q&=&Q\left(\frac{X}{Z},\frac{Y}{Z}\right)=-3\frac{Y}{Z}-2\frac{Y^2}{Z^2}.
\end{eqnarray}
\par\noindent
In terms of the initial variable, Hubble parameter $H$, the asymptotic mapping (\ref{map1}) reads
\begin{equation}\label{map2}
    H^2=\frac{X}{Z},\quad \frac{\dot{H}}{H^2}=\frac{Y}{Z}.
\end{equation}
After a series of manipulations that include time rescaling \cite{Meiss} and the following regularisation of the functions \footnote{Where $m$ is the maximum degree of $P,Q$ here $m=2$.}
\begin{eqnarray}\label{forms}
P^*(X,Y,Z)&=&Z^mP\left(\frac{X}{Z},\frac{Y}{Z}\right)=2XY,\nonumber\\ Q^*(X,Y,Z)&=&Z^mQ\left(\frac{X}{Z},\frac{Y}{Z}\right)=-3YZ(\alpha+1)-2Y^2(\alpha+1),
\end{eqnarray}
\par\noindent
 the singular system (\ref{singular}) becomes complete and valid on $\mathcal{S}^{2+}$.The complete system reads
\begin{eqnarray}\label{completeas}
X'&=&-Y[XQ^*(X,Y)-YP^*(X,Y)]\nonumber\\
Y'&=&X[XQ^*(X,Y)-YP^*(X,Y)].
\end{eqnarray}
\par\noindent
along $X^2+Y^2=1$. It follows that the  equilibrium points at infinity  occurs for
\begin{equation}\label{flow}
XQ^*(X,Y)-YP^*(X,Y)=0,
\end{equation}
\par\noindent
along $X^2+Y^2=1$. 
\par\noindent 
Having regard to the above, the flow defined by system (\ref{asym1}) near its finite-time singularities is equivalent to the flow of the  system (\ref{completeas}) near its equilibria  on the equator of Poincar\`{e} sphere.
\par
Now, substituting the forms (\ref{forms}) into Eq. (\ref{flow}), we come to the conclusion that the equilibria at infinity correspond to those $(X,Y)$ points that satisfy the following condition
\begin{equation}\label{equilibria}
-2XY^2(\alpha+2)=0\quad\text{along}\quad X^2+Y^2=1.
\end{equation}
\par\noindent
These are the points $(X,Y)=(1,0)$ and $(X,Y)=(0,\pm1)$. We note here that the point $(-1,0)$ on Poincar\`{e} disk is not taken into account as it represents the case of contracting universes, already excluded from the analysis. The flow on the equator of Poincar\`{e} sphere is clockwise for those $(X,Y)$ points satisfying $X>0$ \cite{Perko}.

\section{Flow at infinity}\label{infinity}
Let us now study the flow near the equilibria  along the circle of infinity by defining a fan-out map \cite{skot1} for equilibria occurring at $Y>0$ and $X>0$ respectively. 
\par
For example we consider first the equilibrium point $(X,Y)=(0,1)$. Then the transformation 
\begin{equation}
\xi=\frac{X}{Y},\quad \zeta=\frac{Z}{Y}
\end{equation}
\par\noindent
projects $(X,Y,Z)$ onto the plane tangent to the $\mathcal{S}^{2+}$ sphere at $Y=1$. 
\paragraph{Equilibrium on the equator of the Poincar\`{e} sphere satisfying $Y>0$}\label{bb}
\par\noindent
\begin{eqnarray}
\dot{\xi}&=&\zeta^mP^*\left(\frac{\xi}{\zeta},\frac{1}{\zeta}\right)-\xi\zeta^mQ^*\left(\frac{\xi}{\zeta},\frac{1}{\zeta}\right)\\
\dot{\zeta}&=&-\zeta^{m+1}Q^*\left(\frac{\xi}{\zeta},\frac{1}{\zeta}\right).
\end{eqnarray}
\par\noindent
After some manipulations we find that the system reads
\begin{eqnarray}\label{hyber}
\dot{\xi}&=&2\xi(\alpha+1)\nonumber\\
\dot{\zeta}&=&2\zeta(\alpha+1).
\end{eqnarray}
\par\noindent
Note that the equilibrium is at $(\xi_0,\zeta_0)=(0,0)$. The linearization about the origin is 
\begin{equation}\label{inf1}
\mathcal{D}f=
 \left( \begin{array}{cc}
    2(\alpha+2)&0 \\
    0&2(\alpha+1)
  \end{array}\right).
  \end{equation}
  \par\noindent
 The eigenvalues are given by $\lambda_1=2(\alpha+2)$ and $\lambda_2=2(\alpha+1)$. Consequently, the equilibrium point  $(\xi_0,\zeta_0)=(0,0)$ is a hyperbolic point-unstable node since $0<\alpha\leq1$ (see \cite{Meiss} for more information on definition). Thus the flow for the diametrically opposed points with $Y<0$ is opposite. This makes the point $Y=-1$ on the circle of infinity a stable node and an attractor solution at early times. Since
 \begin{equation}
 v=\frac{Y}{Z},\quad\text{and}\quad v=\frac{\dot{H}}{H^2},
 \end{equation}
\par\noindent it occurs that on the circle of infinity $\dot{H}/H^2\sim -1$. After integration, the asymptotic solution for the Hubble parameter satisfies the form $H\sim t^{-1}$, which is diverging on approach to the finite-time singularity. On the other hand, after some manipulation we conclude that  the scale factor satisfies a power-law form solution, approaching zero in the vicinity of the finite-time singularity. Therefore we can conclude that on the circle of infinity, the point $Y=-1$ represents asymptotically the CDM era at early times.
 \paragraph{Equilibrium on the equator of the Poincar\`{e} sphere satisfying $X>0$}\label{dd}
\par\noindent
  To continue with, let us now consider  the equilibrium point $(X,Y)=(1,0)$. Then the transformation 
\begin{equation}
n=\frac{Y}{X},\quad \zeta=\frac{Z}{X}
\end{equation}
\par\noindent
projects $(X,Y,Z)$ onto the plane tangent to the $\mathcal{S}^{2+}$ sphere at $X=1$. 

\begin{eqnarray}
\dot{n}&=&\zeta^mQ^*\left(\frac{1}{\zeta},\frac{n}{\zeta}\right)-n\zeta^mP^*\left(\frac{1}{\zeta},\frac{n}{\zeta}\right)\\
\dot{\zeta}&=&-\zeta^{m+1}P^*\left(\frac{1}{\zeta},\frac{n}{\zeta}\right).
\end{eqnarray}
\par\noindent
After some manipulations we find that the system reads
\begin{eqnarray}\label{nonhyber}
\dot{n}&=&-2n^2(\alpha+2)\nonumber\\
\dot{\zeta}&=&2 n\zeta.
\end{eqnarray}
\par\noindent
Note that the equilibrium is at $(n_0,\zeta_0)=(0,0)$. The linearization about the origin is \begin{equation}\label{inf2}
\mathcal{D}f=
  \left(\begin{array}{cc}
    -4n_0(\alpha+2)&0 \\
    -2\zeta_0&-2n_0
  \end{array}\right),
  \end{equation}
  \par\noindent
  which leads to the doubly degenerate equilibrium case \cite{Meiss} with $\lambda_1=\lambda_2=0$ and $\mathcal{D}f(n_0,\zeta_0)=0$. In order to study such a case of nonhyberbolic node, we transform the system (\ref{nonhyber}) to polar coordinates  by assuming that
  \begin{equation}
      n=r\cos{\theta}\quad\text{and}\quad\zeta=r\sin{\theta}.
  \end{equation}
  \par\noindent
After some manipulation, we derive the following expression
\begin{eqnarray}
\frac{dr}{r}=\frac{-(1+\cos^2{\theta})}{\cos{\theta}\sin{\theta}}d{\theta},
\end{eqnarray}
\par\noindent
which gives (see \cite{Meiss})
\begin{equation}\label{integral}
\ln{r}=\ln{r_0}+\int_{\theta_0}^{\theta} \frac{-2(1+\cos^2{\phi})}{\sin{2\phi}}d{\phi}.
\end{equation}
\par\noindent
From the integral (\ref{integral}) given above, we deduce that the denominator vanishes for $\theta=k\pi$ where $k\in\mathcal{N}^+_0$. These angles $\theta=\theta_c$ define an asymptotic direction of approach to the origin. Hence, the origin is  a nonhyberbolic node and the phase portrait is divided into hyperbolic sectors \cite{Meiss}.
\par\noindent
In terms of the basic variable, the equilibrium point $(1,0)$ corresponds to 
 \begin{equation}
 v=\frac{Y}{Z},\quad\text{thus}\quad v=\frac{\dot{H}}{H^2}\sim0.
 \end{equation}
\par\noindent  After some calculations, the asymptotic solution for the Hubble parameter equals a constant (that is the cosmological constant), while that scale factor satisfies an exponential expansion.  Therefore we can conclude that on the circle of infinity, the point $X=1$ represents asymptotically the de-Sitter universe at late times. The latter is already proven in previous sections.

\section{Classification of  singularities}\label{cla}
In this section, we are interested in the  complete classification of all possible finite-time singularities that the system of equations (\ref{system}) admits.
\par\noindent 
The geometric spacetime singularities of  Einstein equations as predicted by an application of the singularity theorems \cite{HAW1,HAW2}, can be related to possible finite-time singularities that a dynamical system approach to these equations may exhibit \cite{PhD}. However, the nature of these dynamical singularities is a more complicated problem to be addressed in this section. For this purpose, we shall use the techniques developed in \cite{od,card,klaou,sing1,sing2} which are based on the use of an invariant geometric quantity associated to the matter content of the universe, the Bel-Robinson energy.
 \par\noindent 
 For a flat FRW universe, the asymptotic behaviour of the Hubble parameter, the scale factor and the matter fields contribution (electric parts of Bel-Robinson energy) on approach to the finite time singularity ($t\rightarrow0$) provides a complete classification on the dynamical character of the singularity. 
\par
We start our analysis by examining the power-law solution ({\ref{sol1}}). Here the most dominant part of the dynamical quantities read
\begin{equation}
    x(t)=H(t)=\frac{2}{3}t^{-1}, \quad a(t)=t^{2/3}.
\end{equation}
The electric parts of Bel-Robinson energy\footnote{We note here that the first component |E| is related to the total energy density $\rho_{tot}$ and pressure $p_{tot}$ of the interacting fluids, thus the Rayhadhuri equation, whereas the |D| component is related to the total energy density of the interacting fluids, thus the Friedman equation} read
\begin{equation}
    |E|^2=3\left(\frac{\ddot{a}}{a}\right)^2, \quad |D|^2=3\left(\frac{\dot{a}}{a}\right)^4,
\end{equation}
hence
\begin{equation}
    |E|\propto t^{-2}, \quad |D|\propto t^{-1}.
\end{equation}
On approach to the finite-time singularity ($t\rightarrow 0$) the asymptotic quantities above become
\begin{equation}
    a(t)\propto t^{2/3}\rightarrow0,\quad
    H(t)\propto t^{-1}\rightarrow \infty,\quad
    |E(t)|\rightarrow\infty,\quad |D(t)|\rightarrow\infty.
\end{equation}
This is the strongest type out of all singularities, described by the triplet $(S_1,N_1,B_1)$ \cite{klaou} or known as a Big-Bang type singularity. This type of singularity occurs during the very  early stages of the universe and it describes asymptotically a model with a collapsing scale factor and diverging Hubble parameter, energy density and pressure. 
\par\noindent For all the reasons described above, the asymptotic solution (\ref{sol1}) resulting from decompositions (\ref{dc1}) and (\ref{dc3}) exhibits a past-collapse singularity placed at early times with dark matter being the dominant matter component. 
\par
Now, the second interesting solution of our analysis has a completely different asymptotic analysis. As we discussed earlier, at late times the dark energy component is increasingly the dominant matter component and the universe is driven towards de-Sitter spacetime with the Hubble parameter and scale factor satisfying
\begin{equation}
    H(t)=\mathcal{C},\quad a(t)=\exp(\mathcal{C}t),
\end{equation}
while the electric parts of Bel-Robinson energy read
\begin{equation}
    |E|\propto E_s,\quad |D|\propto D_s,
\end{equation}
where $E_s,D_s$ nonzero constant values. It follows that asymptotically at late times the solutions  become
\begin{equation}
    a(t)\rightarrow a_s\not=o,\quad
    H(t)\rightarrow \mathcal{C}\quad
    |E(t)|<\infty,\quad |D(t)|<\infty.
\end{equation}
Here, if we only consider the asymptotic characters of the scale factor and the Hubble parameter, the equation above describes a sudden type singularity $(S_3,N_2)$ \cite{klaou}  or Type II \cite{od,od2,od3}. Nevertheless, in addition to the $(S_3,N_2)$ pair,  the finite values of the  electric parts of  Bel-Robinson is an indication that at late times the singularity is avoided, and the de-Sitter space is complete.  

\section{Discussion}\label{last}
\par\noindent
An asymptotic representation of the structure of solutions at infinity is carried out by the method of central projection to complete the asymptotic results of the method of asymptotic splittings performed previously \cite{p3}. This local method characterises the asymptotic properties of the solutions of a given dynamical system in the vicinity of its finite-time singularity. 
\par
In the beginning of our analysis, through a change of variables we transformed the non linear quasi-homogeneous vector field  into a planar polynomial vector field. We continued by  applying  the Poincar\'{e} central projection to the planar vector field in order to treat infinity as the boundary of an unphysical spacetime \cite{skot1}.
\par
It is shown that for such unified model the interaction is  asymptotically vanishing at early times and the contribution of dark energy (as cosmological constant) is negligible. Hence,  the model is indistinguishable from CDM universe. Such a model attains a pole-like \cite{PhD} type of singularity and it is proved in previous works \cite{p1,p2,p3} that such a dominant behaviour is an attractor of all possible asymptotic solutions on approach to the finite-time singularity placed at early times.
\par
We completed the asymptotic analysis by discussing the stability of the finite equilibria of the  system (\ref{system}). We found that the line of equilibria $y=0$, for $x>0$ is an attractor solution for all orbits that begin off in a certain region in  the phase-plane. We note here that the line of equilibria $y=0$ corresponds to the solution $\dot{H}=0$, that is $H=\mathcal{C}$, where $\mathcal{C}$ (a constant of integration) plays the role of a cosmological constant. Therefore, we conclude that as $t\rightarrow\infty$, the cosmological model will approach de Sitter spacetime.  
\par
On the other hand, solutions that are driven towards infinity while reaching a finite-time singularity, approach a stable node on Poicar\'{e} disc . This is precisely the CDM epoch at early times.
\par
Finally we can conclude that interacting vacuum is indeed viable as at early times, in the limit of vanishing interaction, the model reproduces the properties of a matter dominated universe, while at late times the model approaches de Sitter universe as expected from the generalised Chaplygin gas model.

\end{document}